\documentstyle[preprint,aps]{revtex}

\def\L{\Lambda}
\def\l{\lambda}
\def\Ls{\tilde\Lambda}   
\def\ls{\hat\lambda}
\def\ket{\rangle}
\def\ol{\overline}
\def\beq{\begin{equation}}
\def\eeq{\end{equation}}
\def\beqa{\begin{eqnarray}}
\def\eeqa{\end{eqnarray}}
\def\G{SU(m+1)}
\def\H{SU(m)\times U(1)}
\def\SUM{SU(m)}

\def\s{\sigma}
\def\w{\omega}
\def\N{\cal N}
\def\M{\cal M}

\def\nn{\nonumber}

\def\bp{\begin{picture}}
\def\ep{\end{picture}}

\begin{document}

\title
{CYCLIC PERMUTATIONS IN KAZAMA-SUZUKI \\
STRING MODELS}
\author{G. Aldazabal$^{1,2,3}$, I. Allekotte$^4$, E. Andr\'es$^1$}
\author{and}
\author{C. N\'u\~nez$^{2,5}$}

\bigskip
\address
{$^1$~~ Centro At\'omico Bariloche,
8400 S. C. de Bariloche,\\
Comisi\'on Nacional de Energ\'{\i}a At\'omica,\\
 and Instituto Balseiro, Universidad Nacional de Cuyo, Argentina.
}
\address
{$^2$~~Consejo Nacional de Investigaciones Cient\'{\i}ficas
y T\'ecnicas, Argentina \\
}
\address
{$^3$~~ Universidad Aut\'onoma, Madrid, Spain\\
}
\address
{$^4$~~Max-Planck-Institut fur Physik komplexer Systeme \\
Bayreuther Str. 40, Haus 16, 01187 Dresden, Germany \\
}
\address
{$^5$~~Instituto de Astronom\'{\i}a y F\'{\i}sica del Espacio \\
C.C. 67, suc. 28, 1428 Buenos Aires \\
and Universidad de Buenos Aires, Argentina
}

\date{\today}
\maketitle
\begin{abstract}
Moddings by cyclic permutation symmetries are performed in 4-dimensional
strings, built up from $N=2$
coset models of the type $CP_m=SU(m+1)/SU(m)\times U(1)$.
For  some exemplifying cases, the massless chiral and antichiral
states of $E_6$ are computed. The
extent of the equivalence between different
conformal invariant theories which possess equal
chiral rings  is analyzed.

\end{abstract}
\newpage
\section{Introduction}\label{1}
Algebraic constructions of 4 dimensional strings with $N=1$ spacetime
supersymmetry may be achieved by tensoring $N=2$ superconformal
field theories~(see \cite{gepner,ks1} and references therein).
The $U(1)$ current of the $N=2$ algebra provides the
supersymmetry projector. Projection over odd integer charge states is
required for modular invariance. The theory is anomaly free if the total
internal central charge is $c_{int}=9$.

Models constructed in this way possess a large set of invariances. Modding out
by these  symmetries,  new  modular  invariant  theories  can be constructed.
These orbifolds give rise to more acceptable models from the
phenomenological point  of  view,  in  the  sense  that  the number of chiral
fermion
generations is usually  drastically  reduced.    A complete classification of
these models could be useful for the discussion of interesting questions, such
as mirror symmetry or topology change.

In particular, when two or more
internal theories are equal, permutation symmetries are present.
Actually, the first 3-generation (2,2) string model found by Gepner ~
\cite{gep}
is obtained when moddings by $Z_3$ phase and permutation symmetries
are considered.
A systematic study of cyclic permutation symmetries in Gepner's models was
considered  in  ~\cite{cyclic1,cyclic2}  (see  also  \cite{fiqs}),  where  an
interesting method to compute
the massless spectrum of the orbifoldized theories was introduced.

Supersymmetric  string  vacua  can  also   be  constructed  by  orbifoldizing
Landau-Ginzburg (LG) models \cite{martinec,vafa1,lerche}.
A complete classification of LG superpotentials from which
(2,2) string models
 can be constructed, including permutation
symmetries of identical factors, was performed in ~\cite{kreuzer}.
When the LG superpotentials describe Gepner's models both methods
 yield the same number of $E_6$ generations.
However in the case of LG models
 the chiral ring completely characterizes the theory,
whereas the method introduced in
 ~\cite{cyclic1,cyclic2} requires considering
also  non chiral states and allows to extract more information
(for instance, the partition function of the permuted theory can be
explicitly written).

        In this note we study cyclic
permutation symmetries in 4 dimensional string theories built from
more general N=2 superconformal coset models.
In section II we recall the basics of $CP_m$ cosets which are needed in our
study.
In section III we
review the method to consider orbifolds of LG models with respect
to abelian symmetries \cite{{vafa},{vintri},{kreuzer}}
 and apply it to the computation of the
number of $E_6$ generations of string models constructed with tensor
products of cosets of the type $CP_m = SU(m+1)/SU(m)\times U(1)$. We then
extend the method of references
 ~\cite{cyclic1,cyclic2} to include non diagonal modular
invariant couplings among left and right $CP_m$ theories which
do not admit a LG description.
Even if the basic steps follow~\cite{cyclic1,cyclic2}, the concrete computation
demands a careful study of the component theory. In particular the
construction of the cyclic permutation invariant states requires the knowledge
of the exact weights and charges of the fields of the coset. A closed formula
is known for the chiral (antichiral) fields but an each case computation is
needed for the rest of the fields. In section IV we summarize the general
method of references
{}~\cite{cyclic1,cyclic2} and stress the distinct features
of $CP_m$ cosets. We illustrate the procedure of modding out by cyclic
permutation symmetries of $M$
theories ($M$ prime) with two examples. The
particular case $M=2$ is analyzed in section V. Conclusions are presented in
section VI.   The  explicit  construction  of  the characters for $N=2$ coset
models is
briefly reviewed in the Appendix.

\section{$N=2$ Coset Models}
Here we consider the basic ingredients of $CP_m$ coset models
 following the notation
of~\cite{nos1}. For a more detailed discussion see for example~\cite{fiq}.

We refer to the quotient theory $\frac{SU(m+1)_k\times
SO(2m)_1}{SU(m)_{k+1}\times U(1)}$ as $(m,k)$. $\w_i$ and $\hat \w_i$
denote the fundamental
weight vectors of $\G$ and $\SUM$, with $i$ ranging from $0$ to $m$ and $m-1$
respectively. States of the $N=2$ left superconformal algebra (SCA) are
labelled by $|\L,\l,\Ls\ket$, where $\L$ is a weight vector of $\G$ at level
$k$ ($\L=\sum_{i=1}^{m} m_i \w_i$; $0\leq \sum_{i=1}^{m} m_i\leq k$); $\Ls$ is
 a $SO(2m)$ weight at level 1 (so it can only take the values $0$, $v$, $s$,
$\ol s$) and $\l$ is a weight vector of $\H$ at level $k+1$ obtained by
decomposing $|\L\ket\otimes|\Ls\ket$ into irreducible representations of $\H$.

A general field in the $N=2$ quotient theory, denoted by $\Phi^\l_{\L \Ls}$,
is obtained from the decomposition

\beq
G_{\L} V_{\Ls}= \Phi^\l_{\L \Ls} H_{\l}
\eeq

\noindent where $G_{\L}$, $V_{\Ls}$, $H_{\l}$ correspond to fields in the
representations generated from $|\L\ket$, $|\Ls\ket$, $|\l\ket$ respectively.

The conformal weight $h$ and charge $Q$ of $\Phi^\l_{\L \Ls}$ may in
principle be obtained from this expression. In fact, by considering the
$N=2$ superconformal
 algebra associated to the Kac-Moody algebras $\G$, $SO(2m)$ and $\H$
it is easy to obtain the following equations

\beqa
h&=& {\L (\L + 2 \rho_{m+1})- \l(\l+2 \rho_m) \over 2 (k+m+1)}+
{\Ls^2\over 2} + L\\
Q&=& \sum_{l=1}^{2m} \Ls^l - {2\over k+m+1}(\rho_{m+1}-\rho_m)\cdot\l+2L'=
\nn \\
{}~&=& \sum_{l=1}^{2m} \Ls^l - {q\over k+m+1} +2L'
\eeqa

\noindent where we have further decomposed $\l$ into a $\SUM$ weight
$\ls=\sum_{i=1}^{m-1} \hat n_i \hat \w_i$
and a $U(1)$ charge $q$ (corresponding to the
$U(1)$ of $\H$) as:
\beq
\l=\ls +{\w_m\over m}\,q
\eeq
\noindent $\rho_{m+1}$ and $\rho_m$ denote half the sum of the positive roots
of $\G$ and $\SUM$ respectively. $L,L'$ are integers which should be determined
for each field of the theory, even for the primary ones~\cite{gepnercmp}. The
appearance of these integers relies in the fact that an irreducible
representation
of a Kac-Moody algebra $\hat g$ contains several representations of the
original Lie algebra $g$. Every $\hat g$--representation is generated by the
application of Kac-Moody algebra generators $E_n^{\alpha_i}$, $H_n^{\alpha_i}$
to its highest weight. As $\hat g$ contains $g$ as a subalgebra (generated by
$E_0^{\alpha_i}$ and $H_0^{\alpha_i}$) then any representation of $\hat g$ can
be decomposed into representations of $g$. The difference between the
``$g$-piece" of two weigths of the $\hat g$--representation is a root lattice
vector of $g$. For some particular fields of the theory it is possible to
write down a closed formula for $h$ and $Q$. In fact, for the chiral
fields corresponding to $\L=\l$ and $\Ls =0$, $h=Q/2$ is obtained
from the above equations with $L=L'=0$~\cite{gepnercmp} (and similarly for the
antichiral fields with $\L=\l^*$, where the asterisk means conjugate
representation).

In the study of permutation symmetries the exact weights and
charges of all the primary fields
are necessary, i.e.  $L$ and $L'$ must be determined. They are
computed  in some examples in section IV by explicit construction
of the corresponding characters,
 following~\cite{hny}. This is essentially done
by decomposing the characters $\chi^G_{\L}$, $\chi^{SO(d)}_{\Ls}$ into the
characters $\chi^H_{\l}$ of $\H$. The character of the coset
conformal theory is defined
through

\beq
\chi^G_{\L} \, \chi^{SO(d)}_{\Ls}
=\sum_{\l} \, \chi^{N=2}_{\L,\l,\Ls} \, \chi^H_{\l}.
\eeq

 When counting primary states proper external automorphisms of the
affine Kac-Moody algebra must be taken into account. Under the automorphism
$\s$ a state $|\L,\ls,q,\Ls\ket$ changes
to~\cite{gepnerfi,sierra,caihe} $|\s(\L),\s(\ls),\s(q),\s(\Ls)\ket$, where

\beqa
\s(\L)  &=&
(k-\sum_{i=1}^{m} n_i)\w_1 + \sum_{i=2}^{m} n_{i-1} \w_i\nn\\
\s(\ls) &=& (k+1-\sum_{i=1}^{m-1} n_i)\w_1 + \sum_{i=2}^{m-1} n_{i-1} \w_i\\
\s(q) &=& q+k+m+1 \nn\\
\label{sigma} \s(\Ls=(0),(v),(s),(\bar s)) &=& ((v),(0),(\bar s), (s)) \nn
\eeqa

When applying $\s$ simultaneously to left and right movers the resulting
states must be identified so that the fields in the theory form a unitary
representation of the modular group. Further
 identifications are required, depending on the
symmetry properties of the modular invariants ${\N}$, ${\M}$ defined
below ~\cite{nos1}. As the $CP_m$ models do not possess automorphism
fixed points, this field identification poses no further problem (for a
discussion of fixed points in coset models see~\cite{schell,fpp}).

{}Finally, the full partition function of the $CP_m$  $N=2$ superconformal
theory is

\beq
Z=\sum_{\L,\l,\bar\L,\bar\l,\Ls}\, \chi^{N=2}_{\L,\l,\Ls}\,\,
{\cal N}_{\L,\bar\L}\,\,
{\cal M}_{\l,\bar\l}\,\,\chi^{N=2\;*}_{\bar\L,\bar\l,{\bar {\Ls}}}
\eeq

\noindent In the above equation the sum extends over states $(\L,\l)$ and
$(\bar\L,\bar\l)$ satisfying the condition $C(\L,\l)$ of~\cite{gepnerfi},
which for the Neveu-Schwarz (NS)
 sector corresponds to $\L-\l \in M$, the root lattice of
$\G$. {$\cal N$} and {$\cal M$} denote modular invariants for $\G$ and $\H$
respectively.
Throughout the paper, barred (unbarred)
quantities denote right (left) movers.

To make a string theory out of these models the Gepner construction has to be
followed~\cite{gepner}:
each sector (left and right moving) will be a product of spacetime
bosons and fermions, times the product of $r$ internal $N=2$ coset fields, such
that $c_{int}=9$. To obtain $N=1$ spacetime supersymmetry, a projection over
states with odd integer $U(1)$ charge $Q$ must be performed.

It is useful to denote each state in the full theory by a vector
\beq
V=(\Ls_0;q_1,\dots,q_r;\Ls_1,\dots,\Ls_r)
\eeq
where $\Ls_0$ is a $SO(2)$ weight.
We will also need the definition of a scalar product
\beq
V\cdot V'= \sum_{i=0}^r \Ls_i \Ls'_i
-\sum_{i=1}^r {q_i q'_i\over 2 \eta_i (k_i+m_i+1)}
\eeq
and the vectors
\beqa
\beta_0&=&(\bar s; \eta_1,\dots,\eta_r;s_1,\dots,s_r)\nn\\
\beta_i&=&(v;0,\dots,0;0,\dots,v,\dots,0) \hspace{1.5cm}
\mbox{($v$ in the $i$-th. position)}
\eeqa
Here $\eta_i= {1\over 2} m_i(m_i+1)$.

The supersymmetry projection and aligned boundary conditions, Ramond-Ramond
(R-R)
 and NS-NS,
are now accomplished by
\beqa
\label{qentero}Q(\bar V)= 2\beta_0\cdot \bar V= \mbox{odd integer}\\
\label{evenin}2\beta_i\cdot \bar V= \mbox{even integer}
\eeqa
{}For NS states condition~(\ref{qentero}) amounts to integer
internal charge $Q_{int}$. To maintain modular invariance, twisted sectors
must be included. These are given by the condition
\beq
V=\bar V+s\beta_0+\sum_{i=1}^r n_i\beta_i
\eeq
where $s$ and $n_i$ are integers.
Of course a given state is allowed whenever both invariants $\N$ and $\M$ are
nonvanishing and its multiplicity is given by the product of the modular
coefficients.

The heterotic construction is implemented by replacing left spacetime fermions
by internal free bosons with central charge $c=24$ to cancel the bosonic
anomaly. Modular invariance of the theory follows from the isomorphism between
representations of $SO(2)$ and the new gauge group $E_8\times SO(10)$ under
the modular group. With the supersymmetry projection the gauge group is
enlarged from $E_8\times SO(10)$ to $E_8\times E_6$.


As is well known~\cite{gepner,gq,fiqs,nos1}, $CP_m$ coset models have a
discrete $Z_{k+m+1}$ symmetry. New models can be built from the original one
modding by this symmetry; corresponding string
compactifications may have a smaller number of generations.

These moddings are characterized by a vector
\beq
\Gamma=(0;\gamma_1\eta_1,\dots,\gamma_r\eta_r;0,\dots,0)
\eeq
where $\gamma_i$ are integers satisfying
\beq
2\beta_0\cdot \Gamma = - \sum_{i=1}^r {\gamma_i\eta_i\over k_i+m_i+1}
= integer
\eeq

States in the twisted sectors now verify
\beq
V=\bar V+s\beta_0+\sum_{i=1}^r n_i\beta_i+2x\Gamma
\label{twist}
\eeq
and the generalized GSO projection over states with integer $Q_{int}$ is now
given by the conditions (\ref{qentero}) and (\ref{evenin})
and
\beqa
-\Gamma\cdot (2\bar V+2x\Gamma)&=&integer\label{condii}
\eeqa
(see  ~\cite{fiqs}).
The modular invariant partition function is given by
\beq
Z=\frac{1}{DM_i} \sum_{n_i,m_i,x,y,t,s}  Z(s,n_i,x,t,m_i,y)
\label{orbif}
\eeq
with
\beqa
&Z&(s,n_i,x,t,m_i,y)=\nn\\
&=&\sum_V (-1)^{s+t}
e^{-2i\pi V(t\beta_0+\sum_i m_i\beta_i)}
e^{-2i\pi y\Gamma(2V+2x\Gamma+2s\beta_0)}
\chi_V\chi^*_{V+s\beta_0+\sum_i n_i\beta_i+2x\Gamma}
\eeqa

\noindent
Summations over $x$ and $y$ range from 0 to $M-1$, $M$ being the least integer
such that $M\gamma_i\in  Z$ for all $\gamma_i$. The sum over $y$ implements
condition~(\ref{condii}). Again, only states allowed by the invariants
$\M$ and $\N$ are coupled.

\section{Cyclic Permutations in Landau Ginzburg Models}

When considering $CP_m$ coset  models  with  diagonal modular invariants, the
computation of chiral fermion generations can be simplified by exploiting the
known relation to $N=2$ LG models.
We refer the reader to references \cite{martinec,vafa1,lerche}
for the basics of LG
description of
superconformal $N=2$ theories and here we concentrate on the calculation
of the number of chiral generations in the corresponding superstring vacua.
The LG action in superspace is
\begin{equation}
S = \left(\int d^2z d^4\theta~K(\phi_i,\bar\phi_i)\right) + \left (\int
d^2z d^2\theta ~W(\phi_i)+c.c.\right)
\end{equation}
The superpotential W is holomorphic in the chiral fields $\phi_i$ and,
in the fixed point of the renormalization group trajectories
, defines a superconformal
theory with quasihomogeneous potential
\begin{equation}
W(\lambda^{n_i}\phi_i) = \lambda^D W(\phi_i)
\end{equation}
In order to
describe a $N=2$ string vacuum from a LG theory, the GSO conditions on
integer charge states must be imposed. This corresponds to considering
the orbifold by the discrete group transforming the fields as
\begin{equation}
\Phi_i \rightarrow e^{2\pi i n_i/D} \Phi_i
\label{discg}
\end{equation}
where $n_i$ are the weights and D is the order of the superpotential.

The number of fermionic generations contained in this theory can be
computed with the following formula derived in \cite{vafa,vintri},
\begin{equation}
N_{gen} = -{1\over {2 D}}\sum_{p,s=0}^{D-1} (-)^{({c\over 3}-r)(p+s+ps)}
\prod_{pQ_i, sQ_i \in Z} {{n_i - D}\over n_i}
\end{equation}
where the product is taken over all $n_i$ such that $pQ_i \in Z$ and
$sQ_i \in Z$ simultaneously.

In general, LG models possess a larger discrete symmetry group than that
generated by (\ref {discg}).
They have been classified in reference \cite{kreuzer} where string vacua
as orbifolds by discrete abelian symmetries were constructed. In a base
in which all the elements of the symmetry group act diagonally on the
fields it is possible to characterize each element $g$ of a cyclic group $G$
by the order $M$ of $g$, and by an integer $r$-component vector $\gamma =
(\gamma_1,\dots,\gamma_r)$
\beq
g: \phi_i \rightarrow e^{2\pi i\Theta_i^g} \phi_i = e^{2\pi i\gamma_i/M}\phi_i
\eeq
When $det(g)=1$ for all $g \in G$, the orbifold of the LG model is (2,2)
supersymmetric and still has an interpretation as a Calabi-Yau or an
orbifold of a Calabi-Yau manifold compactification.

The number of generations of a theory orbifoldized by $n$ cyclic groups
characterized by $\gamma^{(1)},\dots,\gamma^{(n)}$ is ~\cite{vintri}
\beq
N_{gen} = -{1\over {2\prod_{i=0}^n M_i}}\sum_{p_i,s_i=0}^{M_i-1}\prod_
{{{\scriptstyle j/}\atop\scriptstyle
\sum_i p_i\gamma_j^{(i)}/M_i \in Z} \atop \scriptstyle
\sum_i s_i\gamma_j^{(i)}/M_i \in Z}
\left (1-{D\over {n_j}}\right )
\label{ngen}
\eeq
where $M_0=D, p_0=p, s_0=s$ and $g^{(0)}=(n_1,\dots,n_r)$. This expression
does not include discrete torsion and holds for $({c\over 3}-r)$ even.
A trivial field may be added whenever this condition is not satisfied,
namely $W'=W+\phi_t^2$.

In order to orbifoldize by discrete symmetries using
this formalism, notice that if a generic potential $W$ has a cyclic permutation
symmetry of $M$ fields ($M$ prime), namely
\beq
W(\phi_0,\dots,\phi_{M-1}) = W(\phi_{M-1},\phi_0,\dots,\phi_{M-2})
\eeq
it is possible to perform the following
change of variables
\beq
\psi_j = {\frac{1}{\sqrt N}} \sum_{k=0}^{M-1} e^{-2\pi ikj/M} \phi_k
\eeq
which trades a permutation symmetry of the fields $\phi_k$ into a phase
symmetry of $\psi_k$ (since $\phi_k\rightarrow\phi_{k+1}$  implies
$\psi_j\rightarrow  e^{2\pi ij/M} \psi_j$).  Notice  that
$\phi_k$ and $\psi_k$ have the same weight, and thus the orbifold by $\beta_0$
is not modified by the change of variables.
The number of
generations of the theory can thus be computed with eqn. (\ref {ngen})
choosing
$\gamma_0=(n,\dots,n,n_M,\dots,n_r)$ and
$\gamma^{(1)}=(0,1,2,\dots,M-1,0,\dots,0)$ of order $M$. The $M$
prime condition
implies
\beq
det (g) = e^{2\pi i\sum_{j=0}^{M-1} {j\over M}} = 1
\eeq
Notice that it is not necessary to know the superpotential explicitly in order
to compute $N_{gen}$ in the orbifoldized theory. Knowledge of the weights and
the order of $W$ suffices.

By applying this procedure to all $N=2$ Gepner models we reobtained the results
of reference \cite{cyclic2}.
The same calculation was performed for $CP_m$ Kazama-Suzuki models
that can be described in terms of a LG superpotential. These are the diagonal
$(m,k)_{AA}$ (the subindices denote the $\cal N$ and $\cal M$ modular
invariants
respectively)
which have ~\cite{gepner*,lerche}
\beq
W = \sum_{j_1+2j_2+\dots+mj_m=k+m+1} A_{n_1\dots
n_m}\phi_1^{j_1}\dots\phi_m^{j_m}
\eeq
where the $A_{n_1\dots n_m}$
coefficients are such that each $\phi_i$ has weight $j$.

The results are listed in Table 1 where the corresponding superpotentials are
denoted by D, $(n_1,\dots,n_r)$ and the cyclic permutations considered can be
deduced from the vectors $\gamma$. We have not included models known to be
equivalent at the chiral level because they clearly yield the same results.
 Notice that eqn.(\ref {ngen}) allows to perform the sum
over the $p_i$'s separately in each sector $(s_0,\dots,s_n)$. Thus the
contribution to $N_{gen}$ can be obtained sector by sector.
However, for non-LG models  another method has to be applied. We
describe it in the following sections, where we also emphasize other
by-products of this formulation.

\newpage
{\center{ TABLE I}}

{\center{Number of generations in $CP_m$ models orbifoldized by
cyclic permutation symmetries}}

\vspace{1cm}

\begin{tabular}{|lllr|}
\cline{1-4}
Model & $D,(n_1,\dots,n_r)$ & $M,\gamma$   & $N_{gen}$ \\
\cline{1-4}
$(2,3)_{AA}\times (2,3)_{AA}\times (2,3)_{AA}$
& 6,(1,2,1,2,1,2) & 3,(0,0,1,1,2,2)& 36\\
\hline
$(2,3)_{AA}\times (2,3)_{AA}\times (2,3)_{AA}$ &
6,(1,2,1,2,1,2) & 2,(0,0,1,1,0,0)& 54\\
\hline
$(4,5)_{AA}\times 1_A\times 1_A\times 1_A$ &
 30,(3,6,9,12,10,10,10) & 3,(0,0,0,0,0,1,2)& 0\\
\hline
$(3,4)_{AA}\times 1_A\times 1_A\times 1_A\times 2_A$ &
 24,(3,6,9,8,8,8,6) & 3,(0,0,0,0,1,2,0)& 0\\
\hline
$(3,3)_{AA}\times 1_A\times 1_A\times 1_A\times 5_A$ &
21,(3,6,9,7,7,7,3) & 3,(0,0,0,0,1,2,0)& 0\\
\hline
$(3,4)_{AA}\times 2_A\times 2_A\times 2_A$ &
8,(1,2,3,2,2,2) & 3,(0,0,0,0,1,2)& 40\\
\hline
$(3,4)_{AA}\times 6_{D}\times 6_{D}$ & 8,(1,2,3,2,3,2,3) &
2,(0,0,0,0,0,1,1)& 40\\
\hline
$(3,3)_{AA}\times 12_{D}\times 12_{D}$ & 7,(1,2,3,1,3,1,3) &
2,(0,0,0,0,0,1,1)& 56\\
\hline
$(2,4)_{AA}\times (2,4)_{AA}\times 5_A$ & 7,(1,2,1,2,1) &
2,(0,0,1,1,0)& 51\\
\hline
$(2,6)_{AA}\times (2,6)_{AA}\times 1_A$ & 9,(1,2,1,2,3) &
2,(0,0,1,1,0)& 42\\
\hline
$(2,6)_{AA}\times 1_A\times 1_A\times 1_A\times
1_A\times 1_A$ & 9,(1,2,3,3,3,3,3) & 3,(0,0,0,1,2,0,0)& 60\\
\hline
$(2,6)_{AA}\times 1_A\times 1_A\times 1_A\times
1_A\times 1_A$ & 9,(1,2,3,3,3,3,3) & 5,(0,0,0,1,2,3,4)& 12\\
\hline
$(2,6)_{AA}\times 1_A\times 1_A\times 1_A\times
1_A\times 1_A$ & 9,(1,2,3,3,3,3,3) & 2,(0,0,0,1,0,1,0)& 48\\
\hline
$(2,6)_{AA}\times 1_A\times 1_A\times 1_A\times 4_A$
 & 18,(2,4,6,6,6,3)& 3,(0,0,0,1,2,0)& 60\\
\hline
$(2,4)_{AA}\times 1_A\times 1_A\times 1_A\times 12_A$
 & 42,(6,12,14,14,14,3) & 3,(0,0,0,1,2,0)& 0\\
\hline
$(2,6)_{AA}\times 1_A\times 1_A\times 2_A\times 2_A$
& 36,(4,8,6,15,6,15) & 2,(0,0,0,0,1,1)& 0\\
\hline
$(2,4)_{AA}\times 26_{D}\times 26_{D}$ & 28,(4,8,2,13,2,13)
& 2,(0,0,0,0,1,1)& 102\\
\hline
\end{tabular}

\newpage
\section{Cyclic Permutations in $CP_{\lowercase{m}}$ cosets}
Cyclic permutation symmetries in Gepner models were systematically
considered in
\cite{cyclic1,cyclic2}. In this section we extend the method developed in
these references to  more general
$CP_m$ coset theories.

Assume that there are $M$ identical $CP_m$ blocks in the internal sector of
the 4-D string. We consider a prime number of identical
theories, but the analysis can be
easily extended to any $M$.
Let us introduce a formal projection operator $P$
over identical states of the $N=2$ superconformal algebra (not necessarily
primary states), such that
$P$ acting on a tensor product of states produces a vanishing
result unless all states have equal charges and weights. After dividing by
this permutation symmetry  the following ``character"
can be defined
\beqa
\nn
\chi_{invar}(\tau) &=& (P+(1-P)/M)\chi_{}(\tau)=
{{\chi_{original}}\over {M}}+{{M-1}\over{M}} P\chi_{original} = \\
 &=& {{\chi ^{M}(\tau)}\over {M}}+{{M-1}\over {M}}
\chi(M\tau)
\eeqa
$P\chi$ formally indicates that the traces must be computed by simultaneously
considering the same state in all blocks, such that $P\chi(\tau)=\chi(M\tau)$.
Each of these states is counted once. The term $(1-P)\over {M}$ corresponds to
the case when at least one state in a block is different from the others.
Because this state could belong to any of the $M$ blocks we must divide by $M$
in order to obtain just one full symmetric state. $\chi_{invar}(\tau)$ does not
transform properly under modular transformations and twisted sectors must be
added. The full modular invariant partition function of the projected theory
may finally be written in terms of the character of just one of the
(identical) component theories
\beq
Z_{new}(\tau,\bar\tau)={Z^{M}(\tau,\bar\tau)\over {M}}+{{M-1}\over M}
Z(M\tau,M\bar\tau)+
{{M-1}\over {M}} \sum_{n=0}^{M-1} Z({{\tau+n}\over M},{{\bar\tau+n}\over M}
) \label{znew}
\eeq
Therefore, if fields of the original $CP_m$ theory are known, (\ref{znew})
allows to compute the spectrum in the modded theory. We shall refer to the
states coming from the last term as twisted states. An analysis of this term
shows that it can be interpreted as the partition function of a new $N=2$
superconformal field theory with central charge $\hat c=Mc$
($c$ is the central charge of each one of the identical theories)
and Virasoro
generators given in terms of those of the original theory by

\beq
{\hat L_m}={L_{mM}\over{M}}+{c (M^{2}-1)\over{24 M}}\delta_{0,m}
\eeq
\beq
{\hat G}_{r}^{\pm}={1\over{\sqrt{M}}}G_{rM}^{\pm}
\eeq
\beq
{\hat J_m}=J_{mM}
\eeq
Therefore the weights and charges of the (twisted) primary states of the new
theory are obtained from the original ones as
\beq
h_{new}={{h+m}\over{M}}+{c (M^{2}-1)\over{24 M} }\label{hnew}
\eeq
\beq
Q_{new}=Q
\label{qnew}
\eeq
where $m$ is the level of the descendant field. Similar expressions are valid
for the right movers.

The sum over $n$ in~(\ref{znew}) imposes the constraint $h+m-\bar h-\bar
m=0 \quad mod \quad M$.

In order to build up the partition function for the 4-D string the spacetime
sector and the other $r-M$\ \ $CP_m$ blocks must be included. Orbifolds by
$\beta_0$ and $\beta_i$ must then be performed. Orbifolds by commuting phase
symmetries could also be considered.

{}Following the reasoning of ref.~\cite{cyclic2}, the $Z_{new}$ corresponding
to~(\ref{znew}) is constructed in terms of the partition function written
in~(\ref{orbif}). Modular transformations allow to build the twisted sector.
The conclusion is that~(\ref{qentero}), (\ref{evenin})
and~(\ref{condii}) are still valid in the
twisted sector if we define the new $1+2(r-M+1)$ component vector

\beqa
V & = & (\Ls_0;q_1,q_{M+1},\dots,q_r;\Ls_1,\Ls_{M+1},\dots,\Ls_r)\nn\\
\beta_0 & = & (\bar s;M
\eta_1,\eta_{M+1},\dots,\eta_r;s_1,s_{M+1},\dots,s_r)\\
\beta_i & = & (v;0,\dots,0;0,\dots,v,\dots,0) \hspace{1.5cm}\nn
\eeqa
and similarly for the phase vector

\beq
\Gamma=(0;M
\gamma_1\eta_1,\gamma_{M+1}\eta_{M+1},\dots,\gamma_r\eta_r;0,\dots,0)
\eeq
where the $M$ original identical theories have been replaced by just one
in the first block.

Eqn. (\ref{twist}) now means that NS left states may couple to NS right
states if
\beqa
q_1-\bar q_1 & = & m_1(m_1+1)M(s+x \gamma_1 ) \qquad
mod \qquad m_1(m_1+1)M(k_1+m_1+1)\label{cond2}   \\
q_i-\bar q_i & = & m_i(m_i+1)(s+x \gamma_i ) \qquad
mod \qquad m_i(m_i+1)(k_i+m_i+1)
\eeqa
for $i=M+1, \dots , r$.

The issue of field identifications in the modded theory deserves a comment.
In fact, in the new theory, expressed in terms of just one original block, the
orbit under $\s$ has been enlarged up to $M(m_1+1)m_1$. An indication of it is
that if we compute the conformal weight $h$ in~(\ref{hnew}) by just using the
$\s$ transformed state~(\ref{sigma}) the weights in the twisted
sector~(\ref{twist}) will be $h_{new}$+ integer, only after applying
$Mm_1(m_1+1)$ times the transformation $\s$~\cite{caihe}.

Massless chiral-chiral (chiral-antichiral) matter (antimatter) fields satisfy
$h_{int}=Q_{int}/2$ and $\bar h_{int}=\bar Q_{int}/2$ ($\bar h_{int}=-\bar
Q_{int}/2$). The chiral (antichiral) states of the tensor product of models
are obtained as products of chiral (antichiral) states of each coset.
Similarly for the new theory, chiral states have $h_{new}=Q_{new}/2=Q/2$.

We will apply the method discussed above to mod out by cyclic permutations of
$M=3$ blocks in the $(2,1)_{AA}^4$ theory ($c_{int}=6$) and in the
$(2,2)_{AA}^33_A $theory ($c_{int}=9$) and compute the massless chiral
spectrum in both cases (${\bf {56}}$ and ${\bf {27}}$ and $\bar {\bf {27}}$ of
$E_6$ respectively).
We indicate the permuted theories in brackets, i.e.
$\{(2,1)_{AA}^3\}(2,1)_{AA}$ and $\{(2,2)_{AA}^3\}3_A$ respectively.

According to notations of section II, in a general $(2,k)$ theory
\beqa
\L & = & m_1 w_1+ m_2 w_2\\
\l & = & n \hat w_1+{q\over 2} w_2
\label{descomp}
\eeqa
such that a state in the coset may be written as $(m_1, m_2)(n,q)s$
(with $s=0,2,1,-1$ corresponding to $0,v,s,{\bar s}$). The constraints
$C(\L,\l)$  read  in the NS sector
\beqa
m_1+2m_2-q&=&0 \qquad mod \qquad 3 \nonumber \\
4m_1+2 m_2-q-3n&=&0 \qquad mod \qquad 6 \nonumber
\eeqa

\noindent (with $m_1+m_2\le {k}$ and $q\equiv q+m(m+1)(k+m+1)$)
and the weights and
charges are
\beq
h={1\over 12(k+m+1)}[4m_{1}^2+4m_{2}^2+4m_{1}m_{2}+12m_1+12m_2-3n(n+2)-q^2]+
{{s^2}\over 8} + L
\eeq
\beq
Q={-q\over{k+m+1}}+{s \over 2}+2L'.
\eeq
Consider the $(2,1)_{AA}$ coset. For this case, when $M=3$,
the above conditions lead to six
highest weight states, namely
$$
\begin{array}{ccccccccccccccc}
state ~\# &~~ & (m_1,m_2) &~~ & n  &~~ & q &~~& s &~~ & h &~~ & Q &~~
& h_{new}\\
1 &~ & (0,0) &~ & 0 &~ &  0 &~ & 0 &~ & 0 &~ & 0 &~~ & {1\over 6} \\
2 &~ & (0,1) &~ & 0 &~ &  2 &~ & 0 &~ & {1\over 4} &~ &
{-1\over 2}&~ &{1\over 4} \\
3 &~ & (1,0) &~ & 0 &~ & -2 &~ & 0 &~ & {1\over 4} &~&
{1\over 2}&~&{1\over 4} \\
4 &~ & (1,0) &~ & 1 &~ &  1 &~ & 0 &~ & {1\over 8} &~&
{-1\over 4}&~&{5\over 24} \\
5 &~ & (0,1) &~ & 1 &~ & -1 &~ & 0 &~ &
{1\over 8} &~&{1\over 4}&~&{5\over 24} \\
6 &~ & (1,0) &~ & 0 &~ &  4 &~ & 2 & ~& {1\over 2}+L &~~ & 2L'
&~~ & {{1+L}\over 3}
\end{array}
$$

\noindent
and their corresponding $\s$ transformed states~(\ref{sigma}). Note that the
integers L, L' entering the definitions (43) and (44) are unknown only for the
sixth field, as this is the only one which is neither chiral nor antichiral. In
order to
exactly evaluate them we look at the corresponding character (see appendix)
expressed as an expansion in powers of $x=e^{2i\pi \tau}$ and $y=e^{2i\pi z}$.
(Note that only the first non vanishing power of $x$ and $y$ respectively are
necessary to determine these integers).

The first terms of the character for this state are
\beq
\chi^{(1,0)}_{(0,4)}=
x^{1/2}y^{0}[1+(y+y^{-1})x^{1/2}+2x+(y+y^{-1})x^{3/2}+4x^2 \dots ]
\eeq
Therefore, $h=1/2 $ and $Q=0 $. The next power of $x$ is increased in
1/2. This is a reflection of the general fact that the original
$N=2$ superconformal algebra may be split into subalgebras built up by the
application of an
even or an odd  number of currents
$G_{-1/2}^{\pm}$ to the original primary state  (see~\cite{gepner}).
The states obtained from the primary by one of these subalgebras are labeled
with $s=0$, and those obtained by the other, with $s= 2$, respectively.
Since the  $\s$  transformation  interchanges $s=0$ and $s=2$, it is possible
that  a    primary    state  has  $s=2$  as  above  (notice  that  the  state
((1,0)(0,4)2)
is identified with ((0,1)(2,0)0)).
As a check of our computations
it can be shown that the other fields in the $\s$ orbit produce the same
character.

Notice that in the case at hand the integers $L$ and $L'$
could have been guessed
by claiming the duality relation (valid after field identifications)
$(m,k)=(k,m)$~\cite{ks1}
so that $(2,1)_{AA}$ corresponds to the well known minimal model
$2_A$. We have explicitly checked this relation by comparing the
characters of the two models up to several powers of $x$ and $y$.

We now turn our attention
to cyclic permutations. We are interested in computing the number
of chiral-chiral and chiral-antichiral states in the projected theory.
{}From~(\ref{hnew}) and~(\ref{qnew}) we find that chiral (and antichiral)
states
in the twisted sector come from states in the $\s$--orbits of

$$
\begin{array}{ccccccccc}
 state ~\#&~~ & ((m_1,m_2)(n,q)s) &~~ & h &~~ & Q &~~ & h_{new} \\
2& &             ((0,1)(0, 2)0) && {1\over 4} && {-1\over 2} && {1\over 4} \\
3& &             ((1,0)(0,-2)0) && {1\over 4} && {1\over 2} && {1\over 4} \\
7& & G_{-1/2}^{+}((1,0)(0, 4)2) && 1 && 1 && {1\over 2} \\
8& & G_{-1/2}^{-}((1,0)(0, 4)2) && 1 && -1 && {1\over 2} \\
9& & G_{-1/2}^{-}((0,1)(1,-1)0) && {5\over 8} && {-3\over 4} && {3\over 8} \\
10&& G_{-1/2}^{+}((1,0)(1, 1)0) && {5\over 8} && {3\over 4} && {3\over 8}
\end{array}
$$

Attaching the chiral states of the fourth $(2,1)_{AA}$ theory (i.e.
the theory which is not permuted) and coupling left and
right movers according to~(\ref{qentero}),
(\ref{evenin}) and~(\ref{condii}) we obtain the
following chiral-chiral states in the twisted sector.
$$ \begin{array}{ccccc}
\{7\} & 1 & - & \{7\} & 1 \\
\{10\} & 5 & - & \{10\} & 5 \\
\{3\} & 3 & - & \{3\} & 3 \\
\{7\} & 1 & - & \sigma ^{6}(\{3\})  & \sigma ^2(3) \\
\{10\} & 5 & - & \sigma ^9(\{10\}) & \sigma ^3(5) \\
\{3\} & 3 & - & \sigma ^{12}(\{7\}) & \sigma ^4(1)
\end{array}
$$
where the numbers denote the state $\#$ in the tables above.

By adding the states in the untwisted sector we finally obtain a total of 20
states. This coincides with the result found in~\cite{cyclic2} for the
$\{2_A\}^32_A$ model, as expected.

 Let us now discuss the second example $\{(2,2)_{AA}^3\} 3_A$.
 The $(2,2)_{AA}$ coset and the $8_{D}$ minimal model are known to have the
same spectrum~\cite{fiq,8d=22a}, so their Poincar\'e polynomials coincide.
 However the full conformal theories are different in the sense that the
$8_{D}$ theory has 25 primary states of the Virasoro algebra while
$(2,2)_{AA}$ has 20. For example the primaries with $(l,q,s)$ and $(l,-q,s)$
equal to (6,4,0), (6,2,0), (6,0,0), (8,4,0) and (8,0,0) are states of
$8_{D}$ but there do not exist corresponding ones with the same weight and
charge  in $(2,2)_{AA}$. Note that the states with $l=4$ have multiplicity
2 due to the modular invariant. Moreover, the characters of some of the
chiral primary fields in $8_{D}$ and $(2,2)_{AA}$ (with the same conformal
weights) turn out to be different. For example the character of the identity
in $8_{D}$ is


\beqa
\nn\chi_{0,0}&=& 1+x+3x^2+6x^3+x^4(13+y^2+y^{-2})+x^{3/2} (y+y^{-1})+ \\
  &\,& +2x^{5/2} (y+y^{-1})+5x^{7/2}(y+y^{-1})+...
\eeqa
while in $(2,2)_{AA}$ it is
\beqa
\nn\chi^{0,0}_{0,0}&=& 1+x+4x^2+8x^3+x^4[19+2(y^{-2}+y^2)]+x^{3/2}
(y^{-1}+y)+ \\ & & +3x^{5/2} (y^{-1}+y)+7x^{7/2} (y^{-1}+y)+...
\eeqa
indicating that both fields transform differently under the conformal algebra.
States that are neither chiral nor antichiral may contribute to form the
twisted chiral states in the new theory (see~(\ref{hnew}),~(\ref{qnew})) and
because these states are not the same for both theories, the
number of states in the twisted sectors may be expected to differ. However,
this is not
the case in the model we are discussing.

The new chiral (antichiral) states in the twisted sector are
$$
\begin{array}{cccc}
 state &  h  & Q & h_{new} \\
 (2,0)(0,\pm 4)0 & {2 \over 5} & \mp {4 \over 5} & ~~{2\over 5}~~ \\
 ~~G^{\pm}_{-1/2}((0,1)(1,5)2)~~ & ~~{7\over 10}~~ & ~~\pm 1~~ & {1\over 2}
\end{array}
$$
Attaching the chiral (antichiral) states of $3_A$ such that the total weight
and charge add up to 1/2 and 1 (-1) we obtain 4 states in the twisted sector
of the {\bf 27} and also of the ${\overline {\bf 27}}$. The same number of
twisted states was obtained in~\cite{cyclic2} for the $\{8_{D}^3\} 3_A$
model.
This coincidence was not expected a priori since the length of the
$\s$ orbits
of both theories are different. However since the left and right movers have
to couple in the $(2,2)_{AA}$ model according to the ${\N}_{\L,\bar\L}$ and
${\M}_{\l,\bar\l}$ modular invariants, only 3 states in the $\s$ chain
potentially contribute to the twisted sector similarly as in the $8_{D}$ case.

In the example we are discussing the equivalent results for both theories can
be explained by noticing that, even if the fields of both models transform
differently under the action of the conformal generators, the partition
functions turn out to be equal. In fact we have explicitly checked that
some characters of the $(2,2)_{AA}$ are sum of characters of the $8_{D}$, and
they
are so in a way that the partition functions of both models coincide. The
remaining characters of the $8_{D}$ and the $(2,2)_{AA}$ models are equal one
to one
(so the corresponding states transform in the same way under the conformal
group).

\section {$Z_2$ permutations}
As discussed in~\cite{cyclic2} care needs to be taken when considering
permutations of one pair of blocks. Since states with $s= 2$
($\bar s= 2$) can be obtained by acting with $G^\pm _{-1/2}$
($\bar G^\pm _{-1/2}$) on the corresponding states with $s=0$ ($\bar s=0$),
permuting a pair of states both having one of these fermionic operators
introduces a minus sign which amounts to the cancellation of the symmetrized
state in the spectrum. Of course this sign will not appear if an even number
of permutations of two theories is considered, the cancellation does not take
place and~(\ref{znew}) directly applies.

The twisted sectors, as shown in~\cite{cyclic2}, have to be constructed in the
R
sector of the original theory in the case of $Z_2$ permutations.

As an example we study $M=2$ permutations in $(2,1)^4$. Consider first two
permutations of two blocks, i.e. $\{(2,1)^2\}\{(2,1)^2\}$. By applying the
spectral flow shift $\beta_0$ to the NS states we obtain
the following set of R states

$$
\begin{array}{cccc}
block~~state & h & h_{new} & Q \\
(0,0)(0,3)1  & 1/16 & 1/8 &  1/4 \\
(0,1)(0,5)1  & 1/16 & 1/8 &  -1/4 \\
(1,0)(0,1)1  & 9/16 & 3/8 &  3/4 \\
(1,0)(1,4)1  & 1/16 & 1/8 &   0  \\
(0,1)(1,2)1  & 5/16 & 1/4 &  1/2 \\
(1,0)(0,7)-1 & 9/16 & 3/8 &  1/4 \\
\end{array}
$$

Coupling left and right movers according to~(\ref{cond2}) we obtain the
following chiral-chiral states in the twisted sector
\def\pa{\{(0,0)(0,3)1\}}
\def\pb{\{(1,0)(0,1)1\}}
\def\pc{\{(0,1)(1,2)1\}}
\beqa
\pa ~ \pb  & - &\pa ~ \pb \nn \\
\pa ~ \pb  & - &\s^6\pa ~ \s^6\pb \nn \\
\pb ~ \pa  & - &\pb ~ \pa \nn \\
\pb ~ \pa  & - &\s^6\pb ~ \s^6\pa \nn \\
\pc ~ \pc  & - &\pc ~ \pc \nn \\
\pc ~ \pc  & - &\s^3\pc ~ \s^3\pc \nn \\
\pc ~ \pc  & - &\s^6\pc ~ \s^6\pc \nn \\
\pc ~ \pc  & - &\s^9\pc ~ \s^9\pc \nn
\eeqa

I.e., a total of eight twisted states. Since this model has four invariant
states, applying~(\ref{znew}) leads to a number of 20 chiral-chiral states
as is consistent with a $c_{int}=6$ model.

Let us now consider the case $\left\{(2,1)^2\right\}~(2,1)^2$, i.e.
permutations of only one pair of blocks. We have to combine the R states
of the ``new'' theory with the NS states of the two unpermuted
theories. Again we obtain eight states in the twisted sector. They are
\def\ea{((1,0)\!(0,\!-2)0)}
\def\eb{((1,\!0)\!(1,\!\!1\!)0)}
\def\ec{((0,0)\!(0, 0)0)}
\def\ed{((0,1)\!(1,-1)0)}
\beqa
\pa\ed\ea &-& \pa\ed\ea \nn \\
\pa\ea\ed &-& \pa\ea\ed \nn \\
\pb\ed\ec &-& \pb\ed\ec \nn \\
\pb\ec\ed &-& \pb\ec\ed \nn \\
\pc\ed\ed &-& \pc\ed\ed \nn \\
\pc\ec\ea &-& \pc\ec\ea \nn \\
\pc\ea\ec &-& \pc\ea\ec \nn \\
\pc\ed\ed &-& \s\!^6\!\pc\s\!^3\!\ed\s\!^3\!\ed \nn
\eeqa

In order to obtain the total number of states in the
${\bf 56}$  representation of $E_6$ we have to take into account
that there is one state in the original theory which cancels when symmetrized.
This leads again to $N_{56}$=20 as expected.

Comparing this example with the equivalent Gepner case, namely
$\left\{2_A^2\right\}~~2_A^2$, a difference arises. In~\cite{cyclic2} the
appearance of the (3,0)-form coupled to the identity is used as a criterion to
decide whether the $Z_2$-permutation is allowed. In the case at hand the state

\beq
(0,0,0)~~(0,0,0)~~-~~(0,-2,-2)~~(0,-2,-2)
\eeq
cancels in the symmetrized combination and thus it is not part of the
spectrum. Notice that if a pair of trivial theories (i.e., $k=0$) were added
and permuted, the (3,0)-form coupled to the identity would not cancel with its
symmetric form.  However in the equivalent Kazama-Suzuki case, i.e.
$\left\{(2,1)^2\right\}~~(2,1)^2$ the (3,0)-form is the state (0,0)(0,-6)0.
Therefore, when coupled to the identity the symmetric combination does not
cancel. This will be the case for all cosets with even $m$.

\section {Summary and Conclusions}
We have computed the number of chiral  generations in the ${\bf 27}$
of $E_6$ for string models
constructed with tensor products of $CP_m$ cosets which  can  be described in
terms of LG superpotentials.
We have also
extended the method developed in~\cite{cyclic1,cyclic2} to study
cyclic permutation symmetries in $CP_m$
string models which do not admit a LG formulation.
This study contributes to a deeper understanding of coset theories. In
particular we used
 explicit formulas (power expansions) for the characters of the
primary fields, based on general constructions of reference~\cite{hny}.
In order to know the exact weights and charges of these fields
just the first power of the characters is needed. Possibly this computation
could be simplified by just isolating this power with some limiting
process~\cite{kac}. Notice that explicit expansions for the characters allow
to check many of the equivalence relations among theories conjectured
essentially from the chiral structures. The information needed from the
original theory in order to construct the projected one, goes beyond the
chiral structure. Therefore theories which are not fully conformally identical
(i.e. their spectra of conformal weights and charges are not equal) may lead
to different results.
For example, in the case of minimal models, if equivalence at the
chiral level is established then the chiral spectrum of the projected theories
coincide. However the spectrum of singlets may differ~\cite{cyclic2}.
An open related question remains, regarding
$Z_2$  permutations  of  $CP_m$  cosets with even $m$. In the example
we have presented (namely $\{(2,1)^2\} (2,1)^2$), they
do  contribute  to  the
massless  spectrum  of  the  orbifoldized  theory, even though the equivalent
Gepner case (i.e. $\{2_A^2\} 2_A^2$) does not.

Moddings by phase symmetries, commuting with the cyclic permutations, may be
easily incorporated into the scheme we have presented. A systematic study of
cyclic permutation  symmetries and phase
moddings in $CP_m$ coset models relevant for string vacua classification will
be published elsewhere.

\section{Acknowledgments}

We are grateful to A. Font and A. Klemm for clarifying comments. This work was
partially supported by Consejo Nacional de Investigaciones Cient\'{\i}ficas y
T\'ecnicas (Argentina); Universidad de Buenos Aires and Fundaci\'on Antorchas.
E.~A. and G.~A. thank Universidad Auton\'oma de Madrid where part of this work
was done.

\section{Appendix}

In this appendix we summarize some useful steps for computing the characters
from~\cite{hny}. There the CP$_2$=$SU(3)_k/(SU(2)_{k+1}\times U(1))$
character is decomposed according to
$$
\frac{SU(3)_k\times SU(2)_1\times U(1)}{SU(2)_{k+1}\times U(1)}=\frac{
SU(3)_k }{SU(2)_k\times U(1)}\times \frac{SU(2)_k\times SU(2)_1}{SU(2)_{k+1}}
\times U(1)
$$
Character formulae for minimal model representations of the second factor
above have been built in~\cite{gko}. The character for a state with charge $m$
of the U(1) theory is given by the $SU(2)_k$ theta function $\theta
_{m,k}(x,y)=\sum_{n\in Z}x^{k\ (n+m/k)^2}y^{k\ (n+m/k)/2} $. The character for
the state $|\L,\l=l~\hat{w}_1+\frac{w_2}{2}q,\Ls=0\ket$ is~\cite{hny}:
$$
\chi _{l,q}^\L (x,y)=\sum_
{\bp(0,0)(20,0)
\put(0,0){\tiny\shortstack{l'=0,...,k \\ m=-5,...,6}}
\ep}
b_{l',m}^\L (x)\ \chi _{l,l'}(x)\ \theta
_{(k+3)\frac{m-q}6+\frac q2;k(k+3)}(x,y)
$$
where b$^\L_{l,q}$ are the branching functions which can be obtained
from~\cite{hny} as

\begin{minipage}{.95\textwidth}
$$
b_{l,q}^\L =\prod_{n=1}^\infty (1-x^{-n})^{-5}\ \sum_
{\bp(1,1)\put(-5,10){\tiny\shortstack{
$\sigma \in W$ }}\ep}
(-1)^{s1+s2}\epsilon (\sigma )
{\bp(1,1)(121,29)\put(0,0){\tiny\shortstack{
$l_1,l_2\in Z $\\$ s_1,s_2=0,...,\infty$
}}\ep}
\times \ x^{\frac 1{2(k+3)}\{[\sigma (\L +\rho )+(k+3)[(l_1+\frac
23s_2-\frac 13s_1)\alpha _1+(l_2+\frac 13s_2+\frac 13s_1)\alpha _2]]^2-2\}}
$$
$$
\times \ x^{-\frac 1{12k}[q+k(s_1+s_2)]^2-\frac
1{4(k+2)}\{[l+1+(k+2)(s_2-s_1)]^2-1\}}
$$
{}~
\end{minipage}

\noindent
where $x=e^{2i\pi \tau }$, $\rho=\alpha _1+\alpha _2$ and $\alpha _1$,$\alpha
_2$ (and $\alpha_3=\alpha _1+\alpha _2$) denote the roots of $SU(3)\,$ with
$\alpha _i^2=2$. The usual normalization $2\frac{\alpha _i\omega _j}{\alpha
_i^2}~=\delta _{ij}$ is taken for the weights. $W$ is the Weyl group of
$SU(3)$. Note that the sum {\it is not} well defined, as it depends on the
order of its terms, the sums over $s_i$ should be performed {\it before} the
sums over $l_i$ to get the right branching function. Doing so, some terms
cancels against others if $p_i>0$ due to the identity~\cite{hny}
$$
\sum_{s=0}^{2p-1}(-1)^sx^{\frac 12(s-(p-\frac 12))^2-\frac 12(p-\frac 12)^2}=0
$$
\noindent
and the sum over $s_i$ for fixed $l_i$ can be rewritten as
$$
\sum_{s=2p}^\infty (-1)^sx^{\frac 12(s-(p-\frac 12))^2-\frac 12(p-\frac12)^2}=
-\sum_{s=0}^\infty (-1)^{-(s+1)}x^{\frac 12(-(s+1)-(p-\frac 12))^2-
\frac12(p-\frac 12)^2}
$$
\noindent
which amounts to the replacement $s \rightarrow -(s+1)$ and a global change of
sign. The following conditions are equivalent to $p>0$ for the $s_i$ sum
$$
\begin{array}{ccc}
s_1: & \qquad  & \frac 13(\alpha _2-\alpha _1)\cdot \sigma (\Lambda +\rho
)+(k+3)(l_2-l_1)+
\frac{l+1}2-\frac q6<0 \\ s_2: & \qquad  & \frac 13(2\alpha _1+\alpha
_2)\cdot \sigma (\Lambda +\rho )+(k+3)\ l_1-\frac{l+1}2-\frac q6<0
\end{array}
$$


\begin{references}

\bibitem{gepner} D. Gepner, Nucl. Phys. {\bf B 296} (1987) 757;
``Lectures on $N=2$ strings", Proceedings of the Trieste Spring School 1989,
M. Green et al. (eds.), Singapore: World Scientific 1990.

\bibitem{ks1} Y. Kazama and H. Suzuki, Nucl. Phys. {\bf B 321} (1989) 232.

\bibitem{gep} D. Gepner, ``String theory on Calabi-Yau manifolds:
the 3 generation case" Princeton preprint (1987)

\bibitem{cyclic1}
A. Klemm, and M. Schmidt, Phys. Lett. {\bf B 245} (1990) 53

\bibitem{cyclic2}
J. Fuchs, A. Klemm, and M. Schmidt, Ann. Phys. {\bf 214} (1992) 221

\bibitem{fiqs} A. Font, L. E. Ib\'a\~nez, F. Quevedo and A. Sierra,
Nucl. Phys. {\bf B 337} (1990) 119.

\bibitem{martinec} E. Martinec, Phys. Lett. {\bf B217} (1989) 431

\bibitem{vafa1} C. Vafa and N. Warner, Phys. Lett. {\bf B218} (1989) 51

\bibitem{lerche} W. Lerche, C. Vafa and N. Warner, Nucl. Phys.
{\bf B 324} (1989) 427

\bibitem{kreuzer}
M.~Kreuzer,R.~Schimmrigk and H.~Skarke, Nucl. Phys. {\bf B 372} (1992) 61.\\
M.~Kreuzer and H.~Skarke, Nucl. Phys. {\bf B 405} (1993) 305.


\bibitem{vafa} C. Vafa, Mod. Phys. Lett. {\bf A 4} (1989) 1169

\bibitem{vintri} K. Intrilligator and C. Vafa, Nucl. Phys. {\bf B 339}
(1990) 95

\bibitem{nos1} G. Aldazabal, I. Allekotte, A. Font and C. N\'u\~nez, Int. J.
Mod. Phys. {\bf A7} (1992) 6273.


\bibitem{fiq}
A. Font, L. Ib\'a\~nez and F. Quevedo, Phys. Lett. {\bf B 217} (1989) 271.


\bibitem{gepnercmp} D. Gepner, Commun. Math. Phys. {\bf 142} (1991) 433.


\bibitem{hny} K.\ Huitu, D.\ Nemeschansky and S.\ Yankielowics,
Phys.\ Lett.\ {\bf B 246} (1990) 105


\bibitem{gepnerfi} D. Gepner, Phys. Lett. {\bf B 222} (1989) 207.

\bibitem{sierra} A. Sierra, ``Field Identification in $N=2$ Cosets",
preprint FTUAM/91-09 (1991).

\bibitem{caihe} Y. Cai and A. He, Phys. Lett. {\bf B 252} (1990) 63.

\bibitem{schell} A. N. Schellekens, Nucl. Phys. {\bf B 366} (1991) 27.

\bibitem{fpp} A. N. Schellekens and S. Yankielowicz, Nucl. Phys. {\bf B 334}
(1990) 67; Int. J. Mod. Phys. {\bf A 5} (1990) 5.

\bibitem{gq} D. Gepner and Z. Qiu, Nucl. Phys. {\bf B 285 [FS19]} (1987) 423.


\bibitem{gepner*} D. Gepner, Nucl. Phys. {\bf B 322} (1989) 65


\bibitem{8d=22a}
B.R.~Greene, C.~Vafa and N.P.~Warner, Nucl. Phys. {\bf B~324} (1989) 371.\\
J.H.~Schwarz, Int. J. Mod. Phys. {\bf A~4} (1989) 2653.

\bibitem{kac} V. G. Kac, ``Infinite-dimensional Lie Algebras--An
Introduction", Birkhauser, Boston, 1983, $2^{nd}$ edition, Cambridge Univ.
Press, Cambridge, 1985.
\bibitem{gko} P.\ Goddard, A.\ Kent and D.\ Olive,
Commun.\ Math.\ Phys. 103 (1986) 105

\end{references}
\end{document}